\documentclass[aps,pre,showpacs,twocolumn,superscriptaddress]{revtex4}
\usepackage{mathrsfs}
\usepackage{amssymb}
\usepackage{amsmath}
\usepackage{graphicx}
\usepackage{dcolumn}
\usepackage{bm}

\begin{document}

\title{Ground-state Riemannian metric, cyclic quantum distance, and the
quantum criticality in an inhomogeneous Ising spin chain}
\author{Yu-Quan Ma}
\affiliation{School of Applied Science, Beijing Information Science and Technology
University, Beijing 100192, China}
\author{Deng-Shan Wang}
\affiliation{School of Applied Science, Beijing Information Science and Technology
University, Beijing 100192, China}
\author{Ya-Jiang Hao}
\affiliation{Department of Physics, University of Science and Technology Beijing, Beijing
100083, China}
\author{Xiang-Guo Yin}
\affiliation{Zentrum f$\ddot{u}$r Optische Quantentechnologien, Universit$\ddot{a}$t
Hamburg, Luruper Chaussee 149, D-22761 Hamburg, Germany}
\author{Wu-Ming Liu}
\affiliation{Beijing National Laboratory for Condensed Matter Physics, Institute of
Physics, Chinese Academy of Sciences, Beijing 100190, China}
\date{\today}

\begin{abstract}
We investigate the ground-state Riemannian metric and the cyclic quantum
distance of an inhomogeneous quantum Ising spin-1/2 chain in a transverse
field. This model can be diagonalized by using a general canonical
transformation to the fermionic Hamiltonian mapped from the spin system. The
ground-state Riemannian metric is derived exactly on a parameter manifold
ring $S^1$, which is introduced by performing a gauge transformation to the
spin Hamiltonian through a twist operator. The ground-state cyclic quantum
distance and the second derivative of the ground-state energy are studied in
different inhomogeneous exchange coupling parameter region. Particularly, we
show that the quantum ferromagnetic phase in the uniform Ising chain can be
characterized by an invariant cyclic quantum distance with a constant
ground-state Riemannian metric, and this metric will rapidly decay to zero
in the paramagnetic phase.
\end{abstract}

\maketitle




\section{Introduction}

Quantum phase transitions (QPTs) are driven purely by the quantum
fluctuations when a parameter of the Hamiltonian describing the system
varies \cite{Sachdev,Sondhi,Vojta}. Traditionally, QPTs can be well
understood in the framework of the Landau-Ginzburg-Wilson paradigm by
resorting to the notions of local order parameter, long range correlations
and symmetry breaking. In the past few years, a lot of efforts have been
devoted into understanding the QPTs from the information-geometry
perspectives \cite{Bengtsson,Ortiz}, such as quantum entanglement \cite%
{ent1,ent2,ent3,ent4}, entanglement entropy \cite{ee1,ee2,ee3,ee4,ee5,ee6},
quantum discord \cite{qd1,qd2,qd3,qd4,qd5}, quantum fidelity and fidelity
susceptibility \cite%
{Zanardi,Venuti,Gu07,Gu10,Chen,Yang,Zhao,Damski1,Dutta,Damski2,Nishiyama},
Berry phase
\cite{Berry,Simon,Pachos,Hamma,Zhu,Ma2009,Hatsugai,Fufu,Ma2012} and
the quantum geometric tensor
\cite{Provost,Berry1989,Resta,Haldane,Ma2010,Rezakhani,Ryu,Neupert,Neupert2}.

Generally, QPTs can be witnessed by some qualitative changes of the
ground-state properties when some parameters of the Hamiltonian across the
quantum critical point (QCP). The underlying physical mechanism lies in the
fact that the different phases are unconnected by the adiabatic evolution of
the ground state. In the vicinity of the QCP, the ground state driven by the
parameters of the Hamiltonian will lead to an avoided energy-level crossing
between the ground state and the first excited state, where the adiabatic
evolution can be destroyed by a vanishing energy gap as the system size
tends to infinity. From perspective of the differential geometry of the
ground state, a monopole as a gapless point in the Hamiltonian parameters
space will generate some interesting effect on the ground-state local or
topological properties, and these properties can be captured by some local
quantities, i.e., the fidelity susceptibility and the Berry curvature; or by
some topological quantum numbers, i.e., the Chern number \cite{TKNN,Niu1984}%
, $Z_2$ number \cite{Kane,Fu,Hasan,Qi}, and recently, the Euler number of
the Bloch states manifold has been proposed \cite{Ma2013,Kolodrubetz}.

Recently, the concept of ground-state quantum geometric tensor has been
introduced to analyze the QPTs. What is surprising is that the two
approaches of the ground-state Berry curvature and the fidelity
susceptibility as a witness to QPTs are unified. Specifically, the real part
of the QGT is a Riemannian metric defined over the parameter manifold, while
the imaginary part is the Berry curvature which flux give rise to the Berry
phase. The Riemannian metric is recognized as the essential part of the
fidelity susceptibility. Generally, the Riemannian metric and the Berry
curvature will exhibit some singularity or scaling behavior in the quantum
critical region under the thermodynamic limit. Particularly, a scaling
analysis of the ground-state quantum geometric tensor in the vicinity of the
critical points has been performed. So far, these approaches have been
applied to detect the phases boundaries in various systems.

In this work, we propose a cyclic quantum distance of the ground state to
detect the QPTs in a transverse field inhomogeneous Ising spin-$1/2 $ chain,
in which the nearest-neighbor exchange interactions will take alternating
parameters between the neighbor sites. This model can be solved exactly by
introducing a general canonical transformation to diagonalize the fermionic
Hamiltonian mapped from the spin Hamiltonian by the Jordan-Wigner
transformation. In our scheme, an extra local gauge transformation is
performed to the spin system by a twist operator, which endows the
Hamiltonian of the system with a topology of a ring $S^1$ without changing
its energy spectrum. We obtain the exact expression of the ground-state
Riemannian metric and study the cyclic quantum distance of the ground state
on the parameter manifold. We study extensively the ground-state Riemannian
metric in different parameter region of the inhomogeneous Ising chain.
Particularly, we show that the quantum ferromagnetic phase in the uniform
Ising chain can be marked by an invariant cyclic quantum distance of the
ground state, and the distance decay to zero rapidly in the paramagnetic
phase.

\section{The model}

Let us consider an inhomogeneous Ising spin-$1/2$ chain, which consists of $N
$ cells with two sites in each cell, and in an external magnetic field. The
Hamiltonian reads
\begin{equation}
\mathcal{H}=-\sum\limits_{l=1}^{N}[J_{a}\sigma _{l,a}^{x}\sigma
_{l,b}^{x}+J_{b}\sigma _{l,b}^{x}\sigma _{l+1,a}^{x}+h(\sigma
_{l,a}^{z}+\sigma _{l,b}^{z})]
\end{equation}%
where $\sigma _{l,m}^{\alpha }(\alpha =x,\ y,\ z;\ m=a,\ b)$ are the local
Pauli operators, $J_{a}\ \left( J_{b}\right) $ is the exchange coupling, $h$
is the external field and the periodic boundary condition (PBC) has been
assumed. A similar inhomogeneous $XY$ spin model has been investigated in
Ref.\cite{Ma2009}, and here we give a brief discussion for the completeness
of this work. First, we subject the system to a local gauge transformation $%
\mathcal{H}(\varphi )=\mathcal{D}_{z}\left( \varphi \right) \mathcal{HD}%
_{z}^{\dag }\left( \varphi \right) $ by a twist operator $\mathcal{D}%
_{z}\left( \varphi \right) =\prod_{l=1}^{N}\exp [i\varphi (\sigma
_{l,a}^{z}+\sigma _{l,b}^{z})/2]$, which in fact makes the system rotate on
the spin along the $z$-direction. It can be verified that $\mathcal{H}%
(\varphi )$ is $\pi $ periodic in the parameter $\varphi $. Considering the
unitarity of the twist operator $\mathcal{D}_{z}\left( \varphi \right) $,
the energy spectrum and critical behavior of the system are obviously
independent with the parameter $\varphi $. The spin Hamiltonian can be
mapped exactly on a spinless fermion model through the Jordan-Wigner
transformation $S_{l,a}^{+}\!=\!C_{l,a}^{\dag }\exp i\pi \!\!\sum_{l^{\prime
}=1}^{l-1}\!\left( C_{l^{\prime },a}^{\dag }C_{l^{\prime },a}+C_{l^{\prime
},b}^{\dag }C_{l^{\prime },b}\right) $ and $S_{l,b}^{+}\!=\!C_{l,b}^{\dag
}\exp i\pi \!\!\sum_{l^{\prime }=1}^{l-1}\!\left( C_{l^{\prime },a}^{\dag
}C_{l^{\prime },a}+C_{l^{\prime },b}^{\dag }C_{l^{\prime },b}\right) +i\pi
C_{l,a}^{\dag }C_{l,a}$, where $S_{l,a(b)}^{\pm }=\left( \sigma
_{l,a(b)}^{x}\pm i\sigma _{l,a(b)}^{x}\right) /2$ are the spin ladder
operators, and $C_{l,a(b)}$ are the fermion operators. The Hamiltonian $%
\mathcal{H}(\varphi )$ is transformed into
\begin{eqnarray}
H(\varphi ) &=&-\sum_{l=1}^{N}J_{a}\left( C_{l,a}^{\dag
}C_{l,b}+e^{-i2\varphi }C_{l,a}^{\dag }C_{l,b}^{\dag }+H.c.\right)   \notag
\\
&&+J_{b}\left( C_{l,b}^{\dag }C_{l+1,a}+e^{-i2\varphi }C_{l,b}^{\dag
}C_{l+1,a}^{\dag }+H.c.\right)   \notag \\
&&+2h(C_{l,a}^{\dag }C_{l,a}+C_{l,b}^{\dag }C_{l,b}-1).
\end{eqnarray}%
Note that the PBC on the spin degrees of freedom $\sigma _{N+1,m}^{\alpha
}=\sigma _{1,m}^{\alpha }$,$(\alpha =x,y,z;\ m=a,b)$ imply that $%
C_{N+1,m}=e^{i\pi N_{F}}C_{1,m}$, where $N_{F}:=\sum_{l^{\prime
}=1}^{N}\sum_{m^{\prime }=a}^{b}C_{l^{\prime },m^{\prime }}^{\dag
}C_{l^{\prime },m^{\prime }}$ denotes the total fermion number. Thus the
boundary conditions on the fermionic system will obey PBC or anti-PBC
depending on whether $N_{F}$ is even or odd. However, the differences
between the two boundary conditions are negligible in the thermodynamic
limit. Without loss of generality, we take the PBC on the fermionic system,
which means $C_{N+1,m}=C_{1,m}$. Second, applying the following Fourier
transformation $C_{l,a}=\frac{1}{\sqrt{N}}\sum_{k}e^{ikR_{la}}\,a_{k}$ and $%
C_{l,b}=\frac{1}{\sqrt{N}}\sum_{k}e^{ik\left( R_{la}+a\right) }\,b_{k}$ to
the Hamiltonian $H(\varphi )$, where $k=(2\pi /2Na)n$, ($n=-\frac{N-1}{2},-%
\frac{N-1}{2}+1,...,\frac{N-1}{2}$and $R_{la}$ ($R_{lb}=R_{la}+a$) is the
coordinate of site $a$ ($b$) on the $l$-th cell in the lattice with the
lattice parameter $2a$. Now, the Hamiltonian $\mathcal{H}(\varphi )$ in the
momentum space reads%
\begin{eqnarray}
H(\varphi )\!\! &=&-\sum_{k}2h(a_{k}^{\dag }a_{k}+b_{k}^{\dag }b_{k}-1)\!
\notag \\
&&+[(J_{a}e^{ika}+J_{b}e^{-ika})a_{k}^{\dag }b_{k}+H.c.]  \notag \\
&&-[(J_{a}e^{i2\varphi +ika}-J_{b}e^{i2\varphi -ika})a_{-k}b_{k}+H.c.]\ .
\end{eqnarray}%
This Hamiltonian $H(\varphi )$ can be exactly diagonalized as%
\begin{equation}
H\left( \varphi \right) =\sum_{q=\gamma ,\eta ,\mu ,\nu }\sum_{k}\Lambda
_{q,k}\left( q_{k}^{\dag }q_{k}-\frac{1}{2}\right) ,  \label{diagham}
\end{equation}%
by using the following canonical transformation, i.e.,
\begin{eqnarray}
\gamma _{k} &=&\frac{1}{\sqrt{2}}(e^{i2\varphi }\cos \frac{\theta _{k}}{2}%
a_{k}+e^{i\delta _{k}}e^{-i\sigma _{k}}\sin \frac{\theta _{k}}{2}%
a_{-k}^{\dag }  \notag \\
&-&e^{i2\varphi }e^{i\delta _{k}}\cos \frac{\theta _{k}}{2}b_{k}+e^{i\sigma
_{k}}\sin \frac{\theta _{k}}{2}b_{-k}^{\dag })\ ,  \notag \\
\eta _{k} &=&\frac{1}{\sqrt{2}}(-e^{-i\delta _{k}}e^{i\sigma {k}}\sin \frac{%
\theta _{k}}{2}a_{k}+e^{-i2\varphi }\cos \frac{\theta _{k}}{2}a_{-k}^{\dag }
\notag \\
&+&e^{i\sigma _{k}}\sin \frac{\theta _{k}}{2}b_{k}+e^{-i2\varphi
}e^{-i\delta _{k}}e^{2i\sigma _{k}}\cos \frac{\theta _{k}}{2}b_{-k}^{\dag
})\ ,  \notag \\
\mu _{k} &=&\frac{1}{\sqrt{2}}(e^{i2\varphi }\cos \frac{\beta _{k}}{2}%
a_{k}-e^{i\delta _{k}}e^{-i\sigma _{k}}\sin \frac{\beta _{k}}{2}a_{-k}^{\dag
}  \notag \\
&+&e^{i2\varphi }e^{i\delta _{k}}\cos \frac{\beta _{k}}{2}b_{k}+e^{i\sigma
_{k}}\sin \frac{\beta _{k}}{2}b_{-k}^{\dag })\ ,  \notag \\
\nu _{k} &=&\frac{1}{\sqrt{2}}(e^{-i\delta _{k}}e^{i\sigma _{k}}\sin \frac{%
\beta _{k}}{2}a_{k}+e^{-i2\varphi }\cos \frac{\beta _{k}}{2}a_{-k}^{\dag }
\notag \\
&+&e^{i\sigma _{k}}\sin \frac{\beta _{k}}{2}b_{k}-e^{-i2\varphi }e^{-i\delta
_{k}}e^{2i\sigma _{k}}\cos \frac{\beta _{k}}{2}b_{-k}^{\dag })\ ,
\label{cantran}
\end{eqnarray}%
where
\begin{eqnarray}
\cos \theta _{k} &=&\frac{2h-M_{k}}{\sqrt{(2h-M_{k})^{2}+N_{k}^{2}}},  \notag
\\
\cos \beta _{k} &=&\frac{2h+M_{k}}{\sqrt{(2h+M_{k})^{2}+N_{k}^{2}}},  \notag
\\
M_{k} &=&\sqrt{J_{a}^{2}+J_{b}^{2}+2J_{a}J_{b}\cos 2ka},  \notag \\
N_{k} &=&\sqrt{J_{a}^{2}+J_{b}^{2}-2J_{a}J_{b}\cos 2ka},  \notag \\
\delta _{k} &=&\arg \left( J_{a}e^{ika}+J_{b}e^{-ika}\right) ,  \notag \\
\sigma _{k} &=&\arg \left( J_{a}e^{ika}-J_{b}e^{-ika}\right) ,
\end{eqnarray}%
and the quasiparticle energy spectrums are
\begin{eqnarray}
\Lambda _{\gamma k} &=&-\frac{1}{2}(2h-M_{k})-\frac{1}{2}\sqrt{%
(2h-M_{k})^{2}+N_{k}^{2}}\ ,  \notag \\
\Lambda _{\eta k} &=&-\frac{1}{2}(2h-M_{k})+\frac{1}{2}\sqrt{%
(2h-M_{k})^{2}+N_{k}^{2}}\ ,  \notag \\
\Lambda _{\mu k} &=&-\frac{1}{2}(2h+M_{k})-\frac{1}{2}\sqrt{%
(2h+M_{k})^{2}+N_{k}^{2}}\ ,  \notag \\
\Lambda _{\nu k} &=&-\frac{1}{2}(2h+M_{k})+\frac{1}{2}\sqrt{%
(2h+M_{k})^{2}+N_{k}^{2}}\ .  \label{egspre}
\end{eqnarray}

\section{Cyclic quantum distance and Ground-state Riemannian metric}

Now, we focus on the geometric properties of the ground state. The
Hamiltonian $H(\varphi )$ in Eq. (\ref{diagham}) has been diagonalized in
the set of quasiparticle number operators, which allows us to determine all
the eigenvalues and eigenvectors. We note that the energy spectrums $\Lambda
_{\eta k}\geq 0$, $\Lambda _{\nu k}\geq 0$ and $\Lambda _{\gamma k}\leq 0$, $%
\Lambda _{\mu k}\leq 0$. The ground state, denoted as $|GS(\varphi )\rangle $%
, corresponds to the state with the lowest energy, which consists of state
with no $\eta $ and $\nu $ fermions occupied but with $\gamma $ and $\mu $
fermions occupied. Explicitly, the ground state can be constructed as
follows
\begin{equation}
\left\vert GS\left( \varphi \right) \right\rangle =\mathcal{N}^{-\frac{1}{2}%
}\prod_{k>0}\left( \gamma _{-k}^{\dag }\gamma _{k}^{\dag }\mu _{-k}^{\dag
}\mu _{k}^{\dag }\eta _{-k}\eta _{k}\nu _{-k}\nu _{k}\right) |0\rangle ,
\label{GS}
\end{equation}%
where $\mathcal{N}^{-\frac{1}{2}}$ is the normalized factor, and $|0\rangle $
are the vacuum states of fermionic operators $a_{k}$ and $b_{k}$,
respectively. It is easy to check that $\eta _{k}|GS(\varphi )\rangle =0$, $%
\nu _{k}|GS(\varphi )\rangle =0$ and $\gamma _{k}^{\dag }|GS(\varphi
)\rangle =0$, $\mu _{k}^{\dag }|GS(\varphi )\rangle =0$ for all $k$. The
corresponding ground-state energy $E_{g}$ is
\begin{eqnarray}
E_{g} &=&\sum_{k}\left( \frac{1}{2}\Lambda _{\gamma k}+\frac{1}{2}\Lambda
_{\mu k}-\frac{1}{2}\Lambda _{\eta k}-\frac{1}{2}\Lambda _{\nu k}\right)
\notag \\
&=&\sum_{k}-\frac{1}{2}\sqrt{(2h-M_{k})^{2}+N_{k}^{2}}  \notag \\
&&-\frac{1}{2}\sqrt{(2h+M_{k})^{2}+N_{k}^{2}}.
\end{eqnarray}

\begin{figure}[t]
\begin{center}
\includegraphics[width=1.65in]{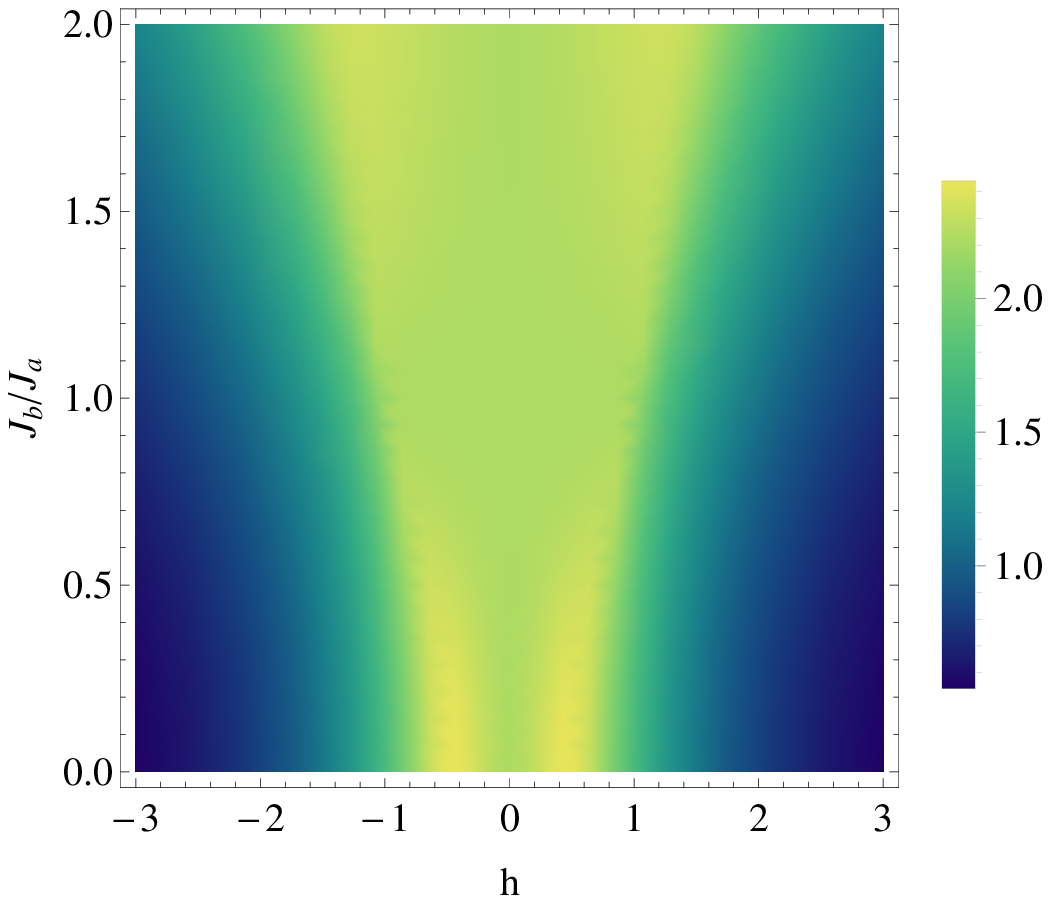} %
\includegraphics[width=1.65in]{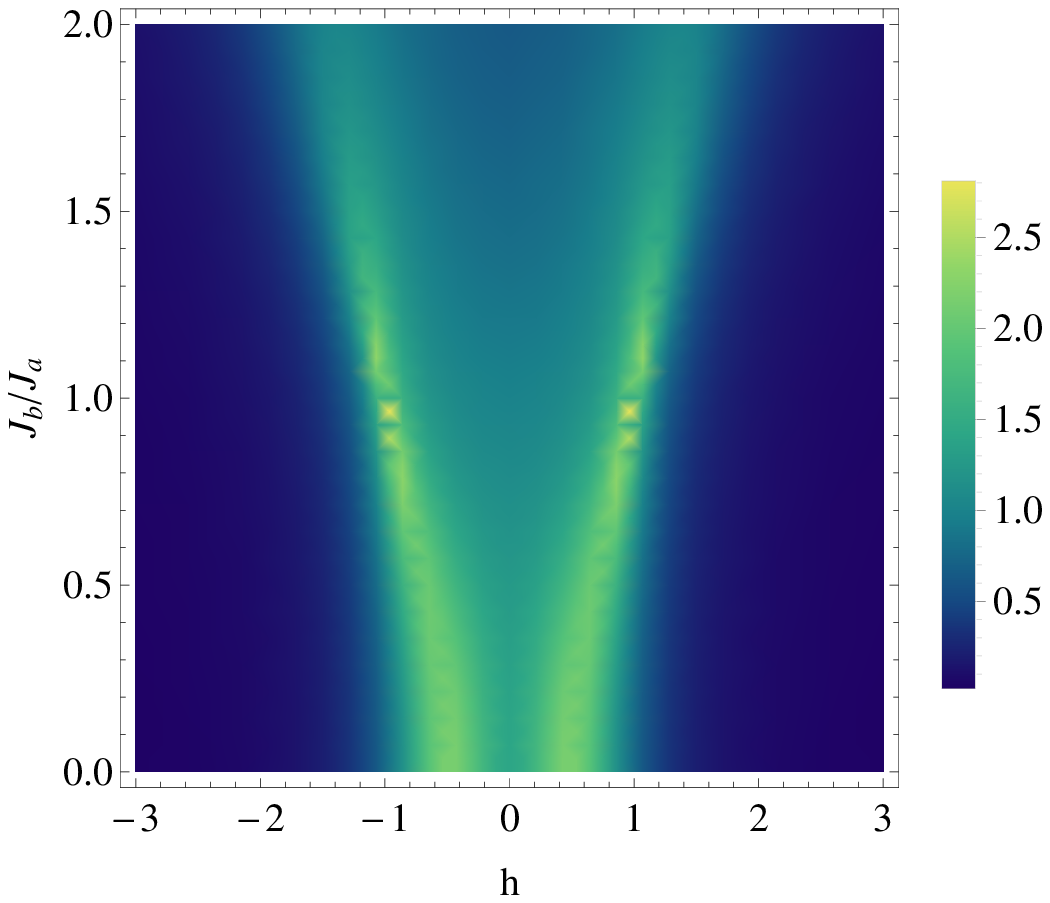}
\end{center}
\caption{(color online) (Left) The cyclical quantum distance $l(0, \protect%
\pi)$ as a function of $J_b/J_a$ and $h$ with the fixed parameters $J_{a}=1$%
, and the system sizes $N\rightarrow \infty$; (Right) The second derivative
of the ground-state energy $E_{g}/N$ with respect to $h$, as a function of $%
J_b/J_a$ and $h$ with the fixed parameters $J_{a}=1$, and the system sizes $%
N=1001$.}
\label{distance3d}
\end{figure}

Now, we introduce the notion of quantum geometric tensor of the ground state
$\left\vert GS\left( \varphi \right) \right\rangle $ on the Hamiltonian
parameter $\varphi $ manifold. It can be verified that the quantum geometric
tensor can be derived from a gauge invariant distance between two ground
states on the $U(1)$ line bundle induced by the quantum adiabatic evolution
of ground states $|g(\varphi )\rangle $ in parameter $\varphi $ space. The
quantum distance $dS$ between two ground states $|GS(\varphi )\rangle $ and $%
|GS(\varphi +\delta \varphi )\rangle $ is given by $dS^{2}=\langle {{{%
\partial _{\varphi }GS(\varphi )d}}\varphi \left\vert {{{\partial _{\varphi
}GS(\varphi )d\varphi }}}\right\rangle }$. Note that the term ${\left\vert {{%
{\partial _{\varphi }GS(\varphi )}}}\right\rangle }$ can be decomposed as $%
\left\vert {{{{\partial _{\varphi }GS(\varphi )}}}}\right\rangle =\left\vert
{D{{{_{\varphi }GS(\varphi )}}}}\right\rangle {+}\left[ \boldsymbol{1}-%
\mathcal{P}({\varphi })\right] \left\vert {{{{\partial _{\varphi }GS(\varphi
)}}}}\right\rangle $, where $\mathcal{P}({\varphi })=\left\vert {GS(\varphi )%
}\right\rangle \left\langle {GS(\varphi )}\right\vert $ is the projection
operator and $\left\vert {D{{{_{\varphi }GS(\varphi )}}}}\right\rangle =%
\mathcal{P}({\varphi })\left\vert {{{{\partial _{\varphi }GS(\varphi )}}}}%
\right\rangle $ is the covariant derivative of $\left\vert {GS(\varphi )}%
\right\rangle $ on the $U(1)$ line bundle. Under the condition of the
quantum adiabatic evolution, the evolution from $\left\vert {GS(\varphi )}%
\right\rangle $ to $|GS(\varphi +\delta \varphi )\rangle $ will undergo a
parallel transport in the sense of Levi-Civit\`{a} from ${\varphi }$ to ${%
\varphi }+\delta {\varphi }$ on the parameter manifold, and hence we have $%
\left\vert {D{{{_{\varphi }GS(\varphi )}}}}\right\rangle =0$. Finally, we
can obtain the quantum distance as $dS^{2}=\langle {{{\partial _{\varphi
}GS(\varphi )}}}|\left( \boldsymbol{1-}\left\vert {GS(\varphi )}%
\right\rangle \left\langle {GS(\varphi )}\right\vert \right) \left\vert {{{%
\partial _{\varphi }GS(\varphi )}}}\right\rangle {{d{\varphi }}}^{2}{.}$The
quantum geometric tensor is given by{\ }%
\begin{equation}
Q_{\varphi \varphi }=\langle {{{\partial _{\varphi }GS(\varphi )}}}|\left(
\boldsymbol{1-}\left\vert GS{(\varphi )}\right\rangle \left\langle {%
GS(\varphi )}\right\vert \right) \left\vert {{{\partial _{\varphi
}GS(\varphi )}}}\right\rangle .  \label{qgt}
\end{equation}%
Substituting Eq. (\ref{GS}) into Eq. (\ref{qgt}), we can derive the concrete
expression of $Q_{\varphi \varphi }$. Obviously, the straightforward
calculation is tedious. We note that the result can be derived concisely
from the following consideration. To begin with, the term $\langle g\left(
\varphi \right) |\partial _{\varphi }g\left( \varphi \right) \rangle $ of $%
Q_{\varphi \varphi }$ can be write as 
\begin{eqnarray}
\langle g\left( \varphi \right) |\partial _{\varphi }g\left( \varphi \right)
\rangle &=&\mathcal{N}^{-1}\prod_{k,j>0}  \notag \\
&&\langle {0}|\left( \nu _{j}^{+}\nu _{-j}^{+}\eta _{j}^{+}\eta _{-j}^{+}\mu
_{j}\mu _{-j}\gamma _{j}\gamma _{-j}\right) \partial _{\varphi }  \notag \\
&&\left( \gamma _{-k}^{\dag }\gamma _{k}^{\dag }\mu _{-k}^{\dag }\mu
_{k}^{\dag }\eta _{-k}\eta _{k}\nu _{-k}\nu _{k}\right) |0\rangle .
\label{term1}
\end{eqnarray}%
%
%
%
%
%
%
%
We note that each term of $\langle {0}|\gamma _{k}\partial _{\varphi }\gamma
_{k}^{\dag }|0\rangle $, $\langle {0}|\gamma _{-k}\partial _{\varphi }\gamma
_{-k}^{\dag }|0\rangle $ in Eq. (\ref{term1}) yield the same results as $%
-2i\cos ^{2}\frac{\theta _{k}}{2}$, and $\langle {0}|\mu _{k}\partial
_{\varphi }\mu _{k}^{\dag }|0\rangle $, $\langle {0}|\mu _{-k}\partial
_{\varphi }\mu _{-k}^{\dag }|0\rangle $ will yield the results as $-2i\cos
^{2}\frac{\beta _{k}}{2}$, meanwhile the other terms yield the results as $0$%
. Finally, we can get
\begin{equation}
\langle {GS}\left( \varphi \right) |\partial _{\varphi }{GS}\left( \varphi
\right) \rangle =\sum_{k>0}-2i\left( \cos ^{2}\frac{\theta _{k}}{2}+\cos ^{2}%
\frac{\beta _{k}}{2}\right) .  \label{Berrycon}
\end{equation}%
Here, we can define a connection as $\mathcal{A}_{\varphi }=i\langle {GS}%
\left( \varphi \right) |\partial _{\varphi }{GS}\left( \varphi \right)
\rangle ,$ which is exactly the Berry-Simon connection of the ground state
on the parameter $\varphi $ manifold. In order to calculate the term of $%
\langle \partial _{\varphi }{GS}\left( \varphi \right) |\partial _{\varphi }{%
GS}\left( \varphi \right) \rangle $, we note that $\langle \partial
_{\varphi }{GS}\left( \varphi \right) |\partial _{\varphi }{GS}\left(
\varphi \right) \rangle =-\langle {GS}\left( \varphi \right) |\partial
_{\varphi }\partial _{\varphi }{GS}\left( \varphi \right) \rangle $ because $%
\partial _{\varphi }\langle {GS}\left( \varphi \right) |\partial _{\varphi }{%
GS}\left( \varphi \right) \rangle =0$ (see Eq. (\ref{Berrycon})), and so we
have
\begin{widetext}
\begin{eqnarray}
\langle \partial _{\varphi }g\left( \varphi \right) |\partial _{\varphi
}g\left( \varphi \right) \rangle  &=&-\mathcal{N}^{-1}\prod_{k,j>0}\langle {0%
}|\left( \nu _{j}^{+}\nu _{-j}^{+}\eta _{j}^{+}\eta _{-j}^{+}\mu _{j}\mu
_{-j}\gamma _{j}\gamma _{-j}\right) \partial _{\varphi }\partial _{\varphi
}\left( \gamma _{-k}^{\dag }\gamma _{k}^{\dag }\mu _{-k}^{\dag }\mu
_{k}^{\dag }\eta _{-k}\eta _{k}\nu _{-k}\nu _{k}\right) |0\rangle   \notag \\
&=&\sum_{k>0}\sum_{j>0}\left( 2i\cos ^{2}\frac{\theta _{k}}{2}+2i\cos ^{2}%
\frac{\beta _{k}}{2}\right) \left( -2i\cos ^{2}\frac{\theta _{j}}{2}-2i\cos
^{2}\frac{\beta _{j}}{2}\right)   \notag \\
&&-\sum_{k>0}4\left( \cos ^{4}\frac{\theta _{k}}{2}+\cos ^{4}\frac{\beta _{k}%
}{2}\right) +\sum_{k>0}4\left( \cos ^{2}\frac{\theta _{k}}{2}+\cos ^{2}\frac{%
\beta _{k}}{2}\right) .  \label{term2}
\end{eqnarray}%
\end{widetext}
Substituting Eq. (\ref{Berrycon}) and Eq. (\ref{term2}) into Eq. (\ref{qgt}%
), we can obtain the QGT as
\begin{eqnarray}
Q_{\varphi \varphi } &=&\langle {{{\partial _{\varphi }g(\varphi )\left\vert
{{{\partial _{\varphi }g(\varphi )}}}\right\rangle -}}}\langle {{{\partial
_{\varphi }g(\varphi )}}}\left\vert {g(\varphi )}\right\rangle \left\langle {%
g(\varphi )}\right\vert {{{\partial _{\varphi }g(\varphi )}}}\rangle  \notag
\\
&=&\sum_{k>0}\sin ^{2}\theta _{k}+\sin ^{2}\beta _{k}.
\end{eqnarray}

Now let us focus on the characterization of the geometric properties of the
ground state. In our approach, the ground state $|{GS}\left( \varphi \right)
\rangle $ is defined in a $U(1)$ line bundle located over a one dimensional
parameter manifold $S^{1}$, and hence the Riemannian metric as the real part
of the QGT is just $Q_{\varphi \varphi }$ itself. As we discussed above, the
ground-state Riemannian metric provide us a gauge invariant distance
measurement of the ground state on the parameter ${\varphi }$ {manifold. }%
The quantum distance $l$ between two ground states $|GS(\varphi _{A})\rangle
$ and $|GS(\varphi _{B})\rangle $ is given by
\begin{equation}
l\left( \varphi _{A},\varphi _{B}\right) =\int_{\varphi _{A}}^{\varphi _{B}}%
\sqrt{\sum\nolimits_{k>0}\sin ^{2}\theta _{k}+\sin ^{2}\beta _{k}}d{\varphi }%
.
\end{equation}%
To have an explicit view of the dependence of the Riemannian metric on the
system size, we can perform a scaling transforming to the $Q_{\varphi
\varphi }$ and denote the Riemannian metric as $g=Q_{\varphi \varphi }/L^{d}=%
\frac{1}{N}\sum_{k>0}\left( \sin ^{2}\theta _{k}+\sin ^{2}\beta _{k}\right) $%
, where $L^{d}=N$ as the number of the sites and here $d=1$ is the dimension
of the system. To study the quantum criticality, we are interested in the
properties under the thermodynamic limit when the system size $N\rightarrow
\infty $, and we have the Riemannian metric
\begin{figure}[t]
\includegraphics[width=3.3in]{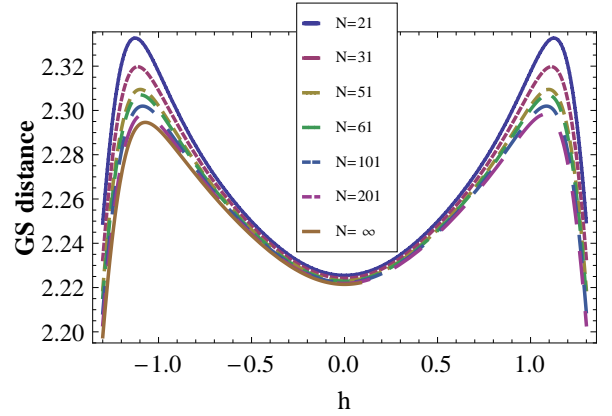}
\caption{(color online) The cyclical quantum distance $l(0,\protect\pi )$ as
a function $h$ with the fixed parameters $J_{a}=1,J_{b}=1.5$, and with
different system sizes.}
\label{distance1.5}
\end{figure}
\begin{figure}[t]
\includegraphics[width=3.3in]{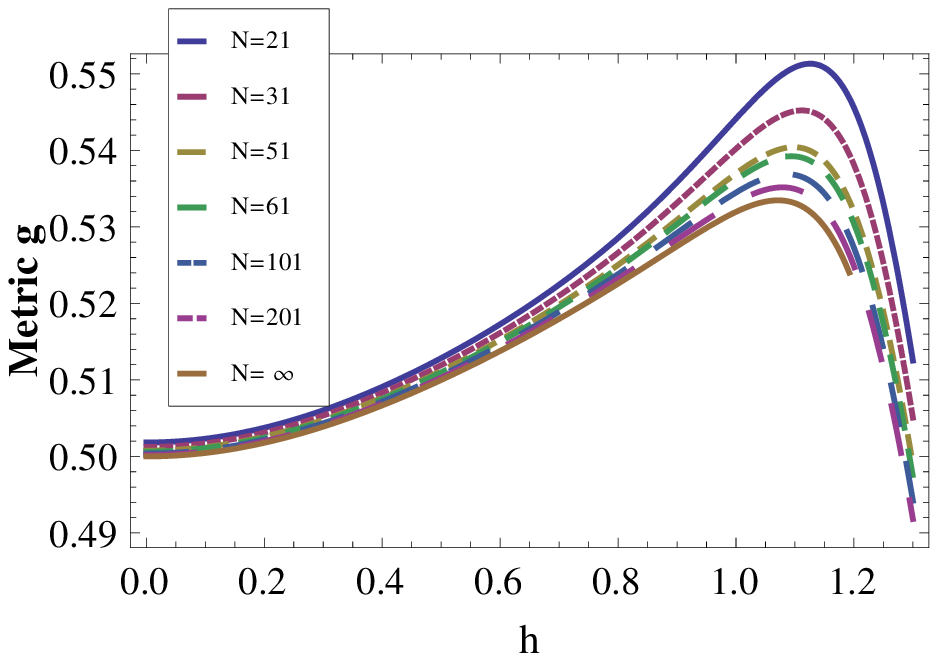}
\caption{(color online) The Riemannian metric $g$ as a function of $h$ with
the fixed parameters $J_{a}=1,J_{b}=1.5$, and with different system sizes.}
\label{metric1.5}
\end{figure}
\begin{figure}[h]
\begin{center}
\includegraphics[width=3.3in]{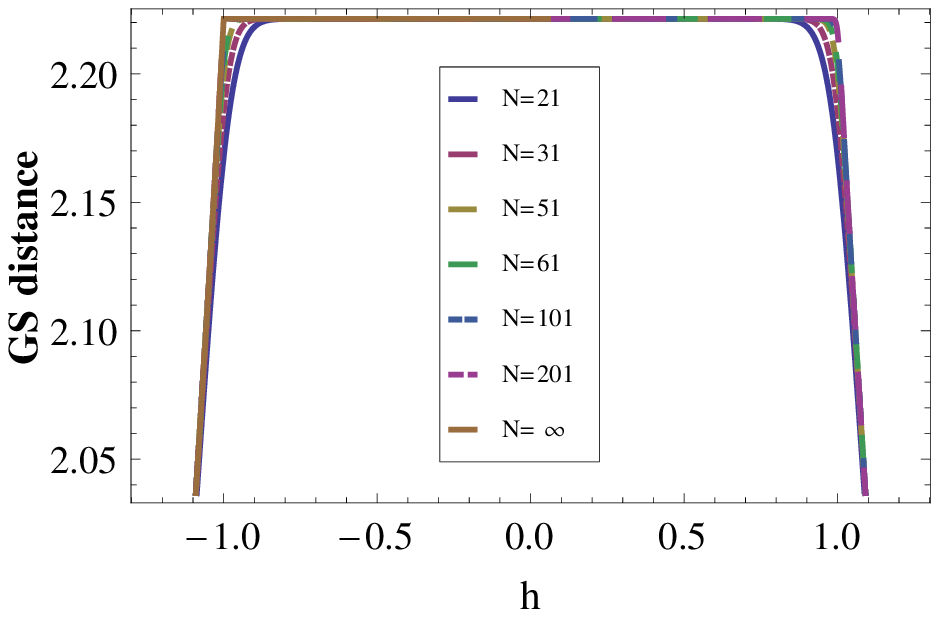}
\end{center}
\caption{(color online) The cyclical quantum distance $l(0,\protect\pi )$ as
a function $h$ with the fixed parameters $J_{a}=J_{a}=1$, and with different
system sizes.}
\label{distance2d}
\end{figure}
\begin{figure}[h]
\includegraphics[width=3.3in]{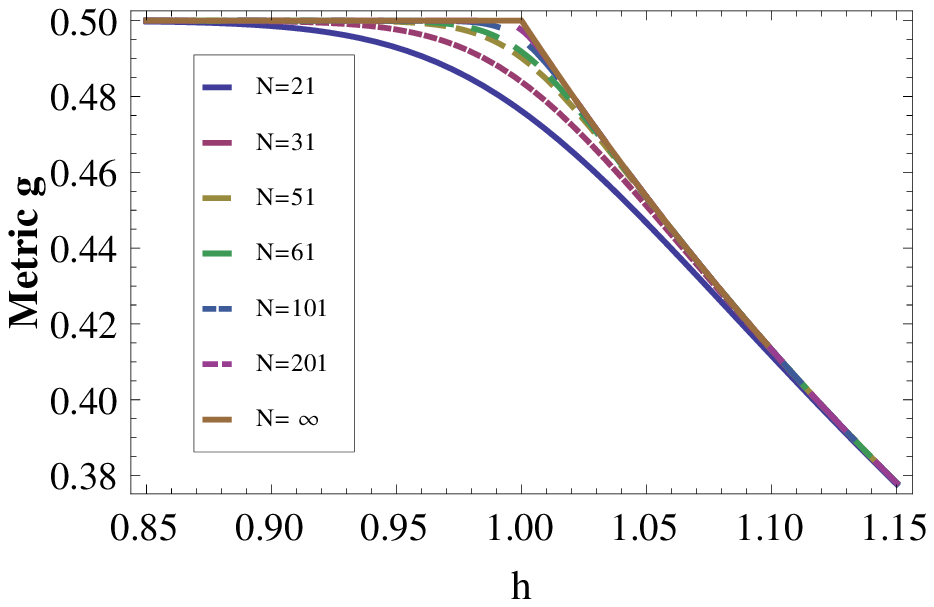}
\caption{(color online) The Riemannian metric $g$ as a function of $h$ with
the fixed parameters $J_{a}=J_{a}=1$, and with different system sizes.}
\label{metric}
\end{figure}
\begin{eqnarray}
g &=&\lim_{N\rightarrow \infty }\frac{1}{N}\sum_{k>0}\left( \sin ^{2}\theta
_{k}+\sin ^{2}\beta _{k}\right)  \notag \\
&=&\frac{1}{2\pi }\int_{0}^{\pi }\frac{\left(
2h^{2}+J_{a}^{2}+J_{b}^{2}\right) \left( J_{a}^{2}+J_{b}^{2}-2J_{a}J_{b}\cos
k\right) }{4h^{4}+\left( J_{a}^{2}+J_{b}^{2}\right)
^{2}-8h^{2}J_{a}J_{b}\cos k}dk,  \label{metric g}
\end{eqnarray}%
where the summation $\frac{1}{N}\sum_{k>0}$ has been replaced by the
integral $\frac{1}{2\pi }\int_{0}^{\pi }dk$. Obviously, the quantum distance
$l\left( 0,\pi \right) $ for a cyclical evolution from $|GS(0)\rangle $ to $%
|GS(\pi )\rangle $ is given by $l\left( 0,\pi \right) =\int_{0}^{\pi }d{%
\varphi }\int_{0}^{\pi }\frac{dk}{2\pi }\left( \sin ^{2}\theta _{k}+\sin
^{2}\beta _{k}\right) $. In Fig. 1(Left), we plot the cyclic quantum
distance $l\left( 0,\pi \right) $ as a function of $h$ and $\alpha
=J_{b}/J_{a}$ with the system size $N\rightarrow \infty $. As a comparison,
we also provide a numeric results of the second derivative of the
ground-state energy with respect to $h$, as a function of $J_{b}/J_{a}$ and $%
h$ with the system sizes $N=1001$ (see Fig. 1(Right)). The cyclic quantum
distance $l\left( 0,\pi \right) $ and the Riemannian metric $g$ as a
function of the external field $h$ with the fixed parameters $J_{a}=1$, $%
J_{b}=1.5$ are shown in Fig. 2 and Fig. 3, respectively. In the region of
inhomogeneous spin exchange coupling $J_{a}\neq J_{b}$, the cyclic quantum
distance and the Riemannian metric has a similar trend with the second
derivative of the ground-state energy. It is worth reminding that, in the
uniform exchange coupling case $J_{a}=J_{b}=1$ and system size $N\rightarrow
\infty $, the ground-state Riemannian metric $g$ (see Eq. (\ref{metric g}))
can be exactly solved as
\begin{eqnarray}
g &=&\frac{1}{2\pi }\int_{0}^{\pi }\frac{\left( h^{2}+1\right) \left( 1-\cos
k\right) }{h^{4}-2h^{2}\cos k+1}dk,  \notag \\
&=&\left\{
\begin{array}{lll}
\frac{1}{2} &  & \text{if}\quad \left\vert h\right\vert \leq 1 \\
\frac{1}{2h^{2}} &  & \text{otherwise}%
\end{array}%
\right. ,
\end{eqnarray}%
which leads to an invariant ground-state cyclical distance $l=\pi /\sqrt{2}$
in the ferromagnetic phase, and $l=\pi /\left( \sqrt{2}h\right) $ in the
paramagnetic phase. In Fig. 4 and Fig. 5, we show the properties of the
cyclic quantum distance and the Riemannian metric in the vicinity of the
critical points with fixed parameters $J_{a}=J_{b}=1$ and different system
size $N$. As shown in Fig. 4 and Fig. 5, the cyclic quantum distance and the
Riemannian metric of the ground state in the ferromagnetic phase is in close
to the constants of $2.22144$ and $0.5$, respectively, with the increase of
the system size $N$. In the thermodynamic limit $N\rightarrow \infty $, the
first derivative of the cyclic distance and the metric are discontinuous in
the critical point.

\section{Conclusion}

In summary, we study the geometric properties of the ground state of a
two-period inhomogeneous quantum Ising chain in a transverse field.
Particularly, we introduce an extra local gauge transformation to the spin
system by a twist operator, which endows the Hamiltonian of the system with
a topology of a ring $S^{1}$ without changing its energy spectrum. On the
parameter manifold, we derive the exact expression of the ground-state
Riemannian metric and define a cyclic quantum distance of the ground state.
We study extensively the ground-state Riemannian metric and the cyclic
quantum distance in different parameter region of the two-period
inhomogeneous Ising chain. Furthermore, we show that the ferromagnetic phase
of a uniform Ising chain can be characterized by a invariant cyclic
ground-state quantum distance, and in the paramagnetic phase the distance
will decay rapidly to zero. This approach provides a interesting description
on the geometric properties of the ground state in addition to the
ground-state Berry phase approach. We hope that the current work will raise
renewed interest in the understanding of the geometric nature of the ground
state in quantum condensed-matter systems.

\section{Acknowledgments}

This work was supported by the special foundation for theoretical
physics research Program of China under grants No. 11347131, the
NKBRSFC under Grants No. 2009CB930701, No. 2010CB922904, No.
2011CB921502, and No. 2012CB821300, NSFC under Grants Nos. 10934010,
11001263, 11004007, and NSFC-RGC under Grants No. 11061160490 and
No. 1386-N-HKU748/10, NSF of Beijing under Grant No. 1132016 and
Beijing Nova program No. xx2013029.

\end{document}